\documentclass[prl,twocolumn,showpacs,superscriptaddress]{revtex4}
\usepackage{graphicx}

\begin{document}

\title{Partonic calculation of the two-photon exchange contribution to
elastic electron-proton scattering at large momentum transfer}
\author{Y.-C. Chen}
\address{Department of Physics, National Taiwan University, 
Taipei 10617, Taiwan}
\author{A. Afanasev}
\address{Thomas Jefferson National Accelerator Facility, 
Newport News, VA 23606, USA}
\author{S. J. Brodsky}
\address{SLAC, Stanford University, Stanford, CA 94309, USA}
\author{C. E. Carlson}
\address{Thomas Jefferson National Accelerator Facility, 
Newport News, VA 23606, USA}
\address{Department of Physics, College of William and Mary,
Williamsburg, VA 23187, USA}
\author{M. Vanderhaeghen}
\address{Thomas Jefferson National Accelerator Facility, 
Newport News, VA 23606, USA}
\address{Department of Physics, College of William and Mary,
Williamsburg, VA 23187, USA}
\date{\today}
\begin{abstract}
We estimate the two-photon exchange contribution to elastic electron-proton 
scattering at large momentum transfer through the scattering
off a parton in the proton. We relate the process on the nucleon 
to the generalized parton distributions which also enter in
other wide angle scattering processes. 
We find that when taking the polarization transfer determinations 
of the form factors as input, adding in the 2 photon correction, 
does reproduce the Rosenbluth data.
\end{abstract}
\maketitle
\begin{figure}
\includegraphics[width=14cm]{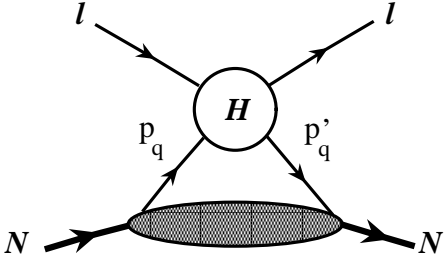}
\vspace{-6.5cm}
\caption{Handbag diagram for the elastic lepton-nucleon
scattering at large momentum transfers. In the hard scattering process $H$, 
the lepton scatters from quarks with momenta $p_q$ and $p'_q$. 
The lower blob represents the GPDs of the nucleon.}
\label{fig:handbag}
\end{figure}

\begin{figure}
\includegraphics[width=11.cm]{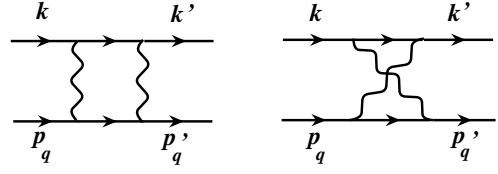}
\vspace{-6.cm}
\caption{Direct and crossed box diagrams to describe the 
two-photon exchange contribution to the lepton-quark scattering
process, corresponding with the blob denoted by $H$ in
Fig.~\ref{fig:handbag}.}
\label{fig:hardbox}
\end{figure}

\begin{figure}
\includegraphics[width=8.cm]{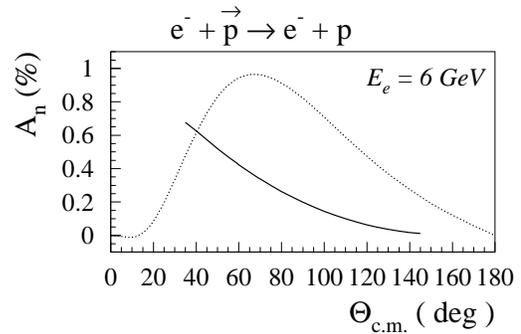}
\vspace{-.6cm}
\caption{Proton normal spin asymmetry  
for elastic $e p$ scattering as function of the {\it c.m.} scattering angle.
The GPD calculation is shown by the solid curve, 
bounded by the kinematic range where $-u, Q^2 > M^2$. 
For comparison, the elastic contribution (nucleon intermediate state in the 
two-photon exchange box diagram) is shown by the dotted curve~\cite{RKR71}.}
\label{fig:an}
\end{figure}

 
%
%
There are currently two experimental methods to extract the 
ratio of electric ($G_{E p}$) to magnetic ($G_{M p}$) proton form factors:
unpolarized measurements employing the Rosenbluth separation technique,
and experiments using a polarized electron beam measuring the ratio of 
the (in-plane) polarization components of the recoiling proton 
parallel and perpendicular to its momentum. 
Recent experiments at Jefferson Lab (JLab) \cite{Chr04,Arr03} 
confirm the earlier Rosenbluth data ~\cite{Slac94}, 
which are at variance with the JLab measurements of $G_{E p}/ G_{M p}$ 
at larger $Q^2$ using the polarization transfer 
technique~\cite{Jones00,Gayou02}. 
This discrepancy casts 
doubt on electron scattering as a precision tool, and needs 
to be sorted out in detail. 
\newline
\indent
Given that no flaws in either experimental technique have been identified
to date, the most likely explanation is that $2 \gamma$ exchange effects 
(beyond those already accounted for in the standard radiative corrections)
are responsible for the discrepancy. The general structure of two- (and
multi)-photon exchange contributions to the elastic electron proton
scattering observables has recently been studied~\cite{GV03}. 
It was found in that work 
that the $2 \gamma$ exchange contribution to the unpolarized
cross section can be kinematically enhanced at larger $Q^2$ 
compared with the $(G_{E p})^2$ term, 
while the $2 \gamma$ exchange contribution to the polarization measurements 
need not affect the results in a significant way. 
A $2 \gamma$ exchange amplitude at the level of a few percent 
may well explain the discrepancy between the two methods. 
To show in a quantitative way 
that $2 \gamma$ exchange effects are indeed able to resolve this 
discrepancy, realistic calculations of  
elastic electron-nucleon scattering beyond the Born approximation are needed. 
One step in this direction was taken recently in~\cite{BMT03}, 
where the contribution to the $2 \gamma$ exchange amplitude was 
calculated for a nucleon intermediate state. This calculation found that the 
$2 \gamma$ exchange correction with the intermediate nucleon has the proper 
sign and magnitude to partially resolve the discrepancy between the two 
experimental techniques. 
In this letter, we report the first calculation of the elastic 
electron-nucleon scattering at large momentum transfer through the scattering
off partons in a nucleon. We relate the process on the nucleon 
to generalized parton distributions (GPDs), which also enter in
other wide angle scattering processes.   
\newline
\indent
To describe the elastic electron-nucleon
scattering~:
\begin{equation}
\label{Eq:intro.2}
l(k)+N(p)\rightarrow l(k')+N(p'),
\end{equation}
we adopt the definitions~:
$P=(p+p')/2$, $K=(k+k')/2$, $q=k-k'=p'-p$,
and choose 
$Q^{2}=-q^{2}$ and $\nu =K \cdot P$ 
as the independent kinematical invariants.  
Neglecting the electron mass, the $T$-matrix for elastic 
electron-nucleon scattering can be expressed through 3 independent 
Lorentz structures as~\cite{GV03}~:
\begin{eqnarray}
\label{eq:tmatrix}
T_{h, \, \lambda'_N \lambda_N} \,&=&\, 
\frac{e^{2}}{Q^{2}} \, \bar{u}(k', h)\gamma _{\mu }u(k, h)\, \\
&&\hspace{-1.75cm} \times \, 
\bar{u}(p', \lambda'_N)\left( \tilde{G}_{M}\, \gamma ^{\mu }
-\tilde{F}_{2}\frac{P^{\mu }}{M}
+\tilde{F}_{3}\frac{\gamma \cdot K 
P^{\mu }}{M^{2}}\right) u(p, \lambda_N), \nonumber
\end{eqnarray}
where $h = \pm 1/2$ is the electron helicity and $\lambda_N$ 
($\lambda'_N$) are
the helicities of the incoming (outgoing) nucleon. 
Furthermore, 
\( \tilde{G}_{M},\, \tilde{F}_{2},\, \tilde{F}_{3} \) are 
complex functions of \( \nu  \) and \( Q^{2} \), and 
we introduced the factor \( e^{2}/Q^{2} \)
for convenience, where \( e\simeq \sqrt{4\pi /137} \) is the proton charge, 
and $M$ is the nucleon mass.   
To separate the one- and two-photon exchange contributions, we 
introduce the decompositions~: $\tilde G_M = G_M + \delta \tilde G_M$, 
and $\tilde F_2 = F_2 + \delta \tilde F_2$, where 
$G_M (F_2)$ are the proton magnetic (Pauli) form factors respectively, 
defined from the matrix element of the electromagnetic current.     
The amplitudes $ \tilde{F}_{3}, \delta \tilde{G}_{M} $
and $\delta \tilde{F}_{2} $ 
originate from processes involving at least the exchange of two photons, 
and are of order $e^2$ 
(relative to the factor \( e^{2} \) in~(\ref{eq:tmatrix})).
The reduced cross section for elastic electron-nucleon scattering, 
including corrections up to order $e^2$ is given by~\cite{GV03}:
\begin{eqnarray}
\sigma_R 
\, &=&\, G_M^2 + \frac{\varepsilon}{\tau}  G_E^2   \,  
+ 2 \, G_M {\cal R} 
\left(\delta \tilde G_M + \varepsilon \frac{\nu}{M^2} \tilde F_3 \right) 
\nonumber \\
&+& 2 \frac{\varepsilon}{\tau} G_E \, 
{\cal R} \left(\delta \tilde G_E + \frac{\nu}{M^2} \tilde F_3 \right) 
+\,  {\mathcal{O}}(e^4), 
\label{eq:crossen} 
\end{eqnarray}
where \( {\cal R} \) stands for the real part,  
$\tau \equiv Q^2 / (4 M^2)$, and 
$\varepsilon \equiv \left(\nu ^{2}-M^{4}\tau (1+\tau ) \right) 
/ \left( \nu ^{2}+M^{4}\tau (1+\tau ) \right)$.
By analogy, we have defined~:
$\tilde{G}_{E}\equiv\tilde{G}_{M}-(1+\tau )\tilde{F}_{2} 
= G_E + \delta \tilde G_E$,    
with $G_{E}$ the proton electric form factor, and $\delta \tilde G_E$ 
the two-photon exchange correction.  
\newline
\indent
An observable which is directly proportional to 
the $2 \gamma$ exchange is given by the elastic scattering of an
unpolarized electron on a proton target polarized normal to the
scattering plane.
The corresponding single spin asymmetry $A_n$  
is related to the absorptive part of the elastic $e N$ scattering
amplitude \cite{RKR71}. 
Since the $1 \gamma$ exchange amplitude 
is purely real, the leading contribution to $A_n$ 
is due to an interference between $1 \gamma$ and $2 \gamma$ exchange. 
It can be expressed at order $O(e^2)$ as (for $m_e = 0$)~:
\begin{eqnarray}
A_n &=& \sqrt{\frac{2 \, \varepsilon \, (1+\varepsilon )}{\tau}} \,\,
\frac{1}{\sigma_R} 
\left\{ - \, G_M \, {\cal I} 
\left(\delta \tilde G_E + \frac{\nu}{M^2} \tilde F_3 \right) \right. 
\nonumber \\
&&\hspace{1cm}\left. + \, G_E \, {\cal I} \left(\delta \tilde G_M 
+ \left( \frac{2 \varepsilon}{1 + \varepsilon} \right) 
\frac{\nu}{M^2} \tilde F_3 \right) \right\} , 
\label{eq:tnsa}
\end{eqnarray}
where $\cal I$ denotes the imaginary part.
\newline
\indent
To estimate the $2 \gamma$ contribution 
to $\delta \tilde G_M$, $\delta \tilde G_E$, and $\tilde F_3$ at large $Q^2$, 
we consider in this letter a partonic calculation as shown on 
Fig.~\ref{fig:handbag}. 
As a first step, we calculate elastic electron-quark scattering 
with massless quarks~: 
$l(k)+q(p_q)\rightarrow l(k')+q(p'_q)$. 
The Mandelstam invariants are given by  
$\hat s \equiv (k + p_q)^2$, $Q^2$, and 
$\hat u \equiv (k - p'_q)^2$, satisfying  
$\hat s + \hat u = Q^2$.
The $T$-matrix for the $2 \gamma$ part of the 
electron-quark hard scattering process can be written as~:
\begin{eqnarray}
\label{eq:tmatrixhard}
H_{h, \, \lambda}^{hard} \,&=&\, 
\frac{e^2}{Q^{2}} \, \bar{u}(k', h)\gamma _{\mu }u(k, h)\,  \\
&\times& \, 
\bar{u} (p'_q, \lambda) \left( e_q^2 \, \tilde{f}_{1} \, \gamma ^{\mu }
\, +\, e_q^2 \, \tilde{f}_{3} \, \gamma  \cdot K \, P_q^{\mu } \, \right) 
u(p_q, \lambda),
\nonumber 
\end{eqnarray}
where $P_q \equiv (p_q + p'_q) / 2$, 
$e_q$ is the fractional quark charge (for a flavor $q$), 
and the quark helicity $\lambda = \pm 1/2$ is
conserved in the hard scattering process.  
For massless quarks, there is no analog of $\tilde{F}_2$ 
in Eq.~(\ref{eq:tmatrix}). 
\newline
\indent
To calculate the hard amplitudes $H_{h, \lambda}^{hard}$, 
we consider the $2 \gamma$ exchange direct and crossed box diagrams of
Fig.~\ref{fig:hardbox}.  
The $2 \gamma$ exchange contribution to the 
elastic electron scattering off elementary spin 1/2 particles has
been calculated before.  
Early references include Refs.~\cite{Nie71,Khr73},  
which we verified explicitly. 
For further use, we separate $\tilde f_1$ 
into a soft and hard part, i.e. 
$\tilde{f}_1 = \tilde{f}_1^{soft} + \tilde{f}_1^{hard}$, using 
the procedure of Ref.~\cite{GY73}.  
The soft part corresponds with the situation where 
one of the photons in Fig.~\ref{fig:hardbox} carries zero four-momentum, 
and is obtained by replacing the other photon's four-momentum by 
$q$ in the numerator and in its propagator in the loop integral. This yields~:
\begin{eqnarray}
{\cal R}\left( \tilde{f}_1^{soft} \right)
&=&\frac{e^2}{4 \pi^2}\,
\left\{ \ln \left( \frac{\lambda^2}{\sqrt{- \hat s \hat u}} \right) 
\ln \left| \frac{\hat s}{\hat u} \right| + \frac{\pi^2}{2} \right\}, 
\label{eq:f1soft} \\
{\cal R}\left( \tilde{f}_1^{hard} \right) 
&=& \frac{e^2}{4 \pi^2} 
\left\{ \, \frac{1}{2} \, \ln \left| \frac{\hat s}{\hat u}  \right| \right.  
\label{eq:f1hard} \\
&+& \left. \frac{Q^2}{4} 
\left[ \frac{1}{\hat u} \ln^2 \left| \frac{\hat s}{Q^2}  \right| 
- \frac{1}{\hat s} \ln^2 \left| \frac{\hat u}{Q^2} \right| 
- \frac{1}{\hat s} \pi^2  \right] \right\}, 
\nonumber 
\end{eqnarray}
where $\tilde f_1^{soft}$, which contains a term 
proportional to $\ln \lambda^2$ ($\lambda$ is 
an infinitesimal photon mass), is IR divergent.  
The real part of $\tilde f_3$ from 
Fig.~\ref{fig:hardbox} is IR finite, and is given by~:
\begin{eqnarray}
{\cal R}\left( \tilde{f}_3 \right)
&=& \frac{e^2}{4 \pi^2} \, \frac{1}{\hat s \, \hat u} \,
\left\{ \hat s \, \ln \left| \frac{\hat s}{Q^2}  \right| + 
\hat u \, \ln \left| \frac{\hat u}{Q^2}  \right| \right. 
\label{eq:f3q} \\
&+&\left. \frac{\hat s - \hat u}{2} 
\left[ \frac{\hat s}{\hat u} \ln^2 \left| \frac{\hat s}{Q^2}  \right|  
- \frac{\hat u}{\hat s} \ln^2 \left| \frac{\hat u}{Q^2}  \right| 
- \frac{\hat u}{\hat s} \pi^2
\right] \right\} . \nonumber
\end{eqnarray}
\newline
\indent
The imaginary parts of $\tilde f_1$ and $\tilde f_3$ 
originate solely from the direct $2 \gamma$ exchange box diagram 
of Fig.~\ref{fig:hardbox} and are~:
\begin{eqnarray}
{\cal I}\left( \tilde{f}_1^{soft} \right)
&=& - \frac{e^2}{4 \pi} \, 
\ln \left( \frac{\lambda^2}{\hat s} \right),  
\label{eq:f1qimsoft} \\
{\cal I}\left( \tilde{f}_1^{hard} \right)
&=& - \frac{e^2}{4 \pi} 
\left\{ \frac{Q^2}{2 \, \hat u}  
\ln \left( \frac{\hat s}{Q^2}  \right) 
+ \frac{1}{2} \right\}, 
\label{eq:f1qimhard} \\
{\cal I}\left( \tilde{f}_3 \right)
&=& - \frac{e^2}{4 \pi} 
\, \frac{1}{\hat u} \, \left\{ 
\frac{\hat s - \hat u}{\hat u}  
\ln \left( \frac{\hat s}{Q^2}  \right) 
+ 1 \right\}.  
\label{eq:f3qim}
\end{eqnarray}
\indent
Having calculated the hard subprocess, 
we next discuss how to embed the quarks in the nucleon. 
We begin by discussing the soft contributions. There are also soft 
contributions from processes where the photons interact with different quarks. 
One can show that the IR contributions from these processes, which are 
proportional to the products of the charges of the interacting quarks, 
added to the soft contributions from the handbag diagrams  
give the same result as the soft contributions calculated with just 
a nucleon intermediate state. Thus the low energy theorem is satisfied. 
The hard parts when the photons couple to different quarks are subleading 
in $Q^2$ because of momentum mismatches in the wavefunctions.  
\newline
\indent
For the real parts, the IR divergence 
arising from the direct and crossed box diagrams 
is cancelled when adding the bremsstrahlung interference 
contribution with a soft photon emitted from the electron and proton. 
This provides a radiative correction term proportional 
to the target charge $Z$, which may be written as~:
\begin{eqnarray}
\label{eq:crosssoft}
\sigma_{R, soft} = \sigma_{1 \gamma} \, 
\left( 1 + \delta_{2 \gamma}^{soft} + \delta_{brems}^{e p} \right), 
\end{eqnarray}
where $\sigma_{1 \gamma}$ is the $1 \gamma$ exchange cross 
section. In~(\ref{eq:crosssoft}), 
the soft-photon part of the nucleon box diagram is given by 
\begin{eqnarray}
&&\delta_{2 \gamma}^{soft} 
= \frac{e^2}{2 \pi^2} \left\{ \ln \left( 
\frac{\lambda^2}{\sqrt{(s - M^2) |u - M^2|}}  \right) \, 
\ln \left| \frac{s - M^2}{u - M^2} \right| \right. 
\nonumber \\
&&\;\;- \, L\left( \frac{s - M^2}{s} \right) 
- \frac{1}{2} \ln^2\left( \frac{s - M^2}{s}\right) \nonumber \\
&&\;\; + \, 
\left. {\cal R} \left[ L\left( \frac{u - M^2}{u} \right) \right] 
+ \frac{1}{2} \ln^2\left( \frac{u - M^2}{u}\right) + \frac{\pi^2}{2} 
\right\} , 
\label{eq:delta2gsoft}
\end{eqnarray}
where $L$ is the Spence function.  
The bremsstrahlung contribution $\delta_{brems}^{ep}$ 
we take from Ref.~\cite{MT00} (see their Eq.~(4.14)).
When comparing with elastic $ep$ cross section data, which are 
usually radiatively corrected using the Mo and Tsai procedure~\cite{MoTsai68}, 
we only have to consider the difference between our above  IR finite 
$\delta_{2 \gamma}^{soft} + \delta_{brems}^{ep}$ and the 
expression for the $Z$-dependent radiative correction in \cite{MoTsai68}. 
This difference is predominantly given by a constant shift proportional to 
$\pi^2 / 2$ in Eq.~(\ref{eq:delta2gsoft}). 
\newline
\indent
We next discuss the hard $2 \gamma$ exchange contribution.
In the kinematical regime where  
$s = (p + k)^2$, $-u = -(p - k')^2$ and 
$Q^2$ are large compared to a hadronic scale ($s, -u, Q^2 >> M^2$), 
this part of the amplitude is calculated as a convolution between an 
electron-quark hard scattering and a soft nucleon matrix element. 
It is convenient to choose 
a frame where $q^+ = 0$, where we introduce light-cone variables 
$a^\pm$  proportional to $(a^0 \pm a^3)$ and choose $P^3 > 0$. 
In the frame $q^+ = 0$, the + momentum fractions of electrons and partons  
are defined as $\eta = K^+ / P^+$ and $x = P_q^{+} / P^+$ respectively.   
At large $Q^2$, we can neglect the intrinsic transverse 
momenta of the active quarks.
The Mandelstam invariants for the hard process are then given by 
$\hat s = Q^2 (x + \eta)^2 / (4 x \eta)$ and 
$\hat u = - Q^2 (x - \eta)^2 / (4 x \eta)$. 
We extend the handbag formalism~\cite{SJB72}, used in 
wide angle Compton scattering~\cite{Rad98,Die99},  
to the $2 \gamma$ exchange process in elastic $e p$ scattering, 
and keep the $x$ dependence in the hard scattering amplitude. 
This yields the $T$-matrix for the process (\ref{Eq:intro.2}) as~:
\begin{eqnarray}
\label{eq:handbag}
T_{h, \, \lambda'_N \lambda_N}^{hard} &=&
 \int_{-1}^1 \frac{dx}{x} \, \sum_q \frac{1}{2}
\left[ H_{h,\, + \frac{1}{2}}^{hard} + H_{h,\,- \frac{1}{2}}^{hard} \right] 
\nonumber \\
&\times& \left[\, H^q\left(x, 0, q^2 \right) \, 
\bar{u}(p', \lambda'_N) \, \gamma \cdot n \, u(p, \lambda_N) \right. 
\nonumber \\
&&\left. \hspace{-0.25cm} +\, E^q\left(x, 0, q^2 \right)  \, 
\bar{u}(p', \lambda'_N) \, \frac{i \, \sigma^{\mu \nu} \, n_\mu q_\nu}{2 M} 
\, u(p, \lambda_N) \right] \nonumber \\
&+&  \int_{-1}^1 \frac{dx}{x} \, \sum_q \frac{1}{2}
\left[ H_{h,\, + \frac{1}{2}}^{hard} - H_{h,\,- \frac{1}{2}}^{hard} 
\right] \, \mathrm{sgn}(x) \nonumber\\ 
&\times& \,
\tilde H^q\left(x, 0, q^2 \right) 
\bar{u}(p', \lambda'_N) \, \gamma \cdot n \, \gamma_5 \, u(p, \lambda_N),
\end{eqnarray}
where $H^{hard}$ is evaluated 
using $\tilde f_1^{hard}$ and $\tilde f_3$, 
and where 
$n^\mu = 2 / \sqrt{M^4 -  s u} 
\left(- \eta \, P^\mu \,+\, K^\mu \right)$
is a Sudakov four-vector ($n^2 = 0$). 
Furthermore, 
$H^q, E^q, \tilde H^q$ are the GPDs for a quark $q$ in the
nucleon. 
\begin{figure}
\includegraphics[width=8.75cm]{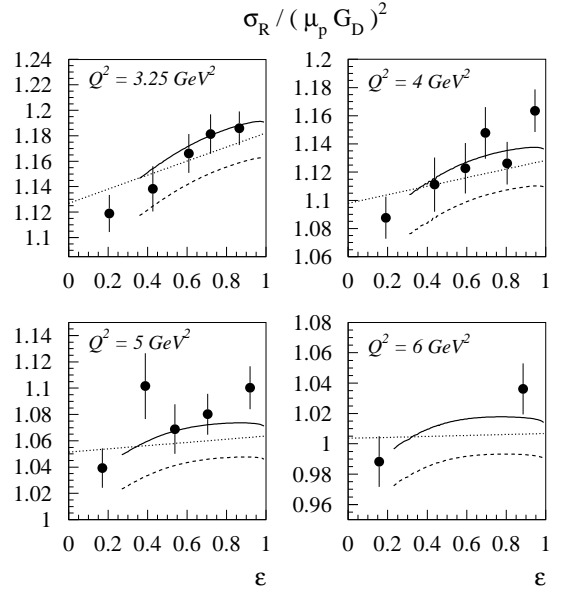}
\vspace{-1.25cm}
\caption{Rosenbluth plots for elastic $e p $ scattering : 
$\sigma_R$ divided by $(\mu_p G_D)^2$, 
with $G_D = (1 + Q^2 / 0.71)^{-2}$.  
Dotted curves : Born approximation using $G_{E p} / G_{M p}$ from 
polarization data~\cite{Jones00,Gayou02}.  
Dashed curves : results when adding the 
GPD calculation for the hard $2 \gamma$ exchange correction, 
for the kinematical range $s, -u > M^2$. 
Full curves are the total results, including 
in addition the soft $2 \gamma$ exchange 
correction relative to Mo and Tsai.  
The data are from Ref.~\cite{Slac94}.}
\label{fig:cross}
\end{figure}
\newline
\indent
From Eqs.~(\ref{eq:tmatrix}),(\ref{eq:tmatrixhard}), and (\ref{eq:handbag})   
the hard $2 \gamma$ exchange contributions to 
$\delta \tilde G_M$, $\delta \tilde G_E, \tilde F_3$ 
are obtained as~:
\begin{eqnarray}
\hspace{-0.4cm}
\delta \tilde{G}_M^{hard} &=& C, 
\label{eq:GMhandbag}\\
\hspace{-0.4cm}
\delta \tilde{G}_E^{hard} &=& 
- \left( \frac{1 + \varepsilon}{2 \varepsilon} \right) \, (A - C) 
+ \sqrt{\frac{1 + \varepsilon}{2 \varepsilon}} \, B , 
\label{eq:GEhandbag} \\
\hspace{-0.4cm}
\tilde F_3 &=& \frac{M^2}{\nu}
\left( \frac{1 + \varepsilon}{2 \varepsilon} \right) \, (A - C ),
\label{eq:F3handbag}
\end{eqnarray}
with
\begin{eqnarray}
A &\equiv& \int_{-1}^1 \frac{dx}{x}     
\frac{\left[(\hat s - \hat u) \tilde{f}_1^{hard} -
\hat s \hat u \tilde{f}_3 \right]}{s - u} 
\sum_q e_q^2 \, \left( H^q + E^q \right), 
\nonumber \\
B &\equiv& \int_{-1}^1 \frac{dx}{x}     
\frac{\left[(\hat s - \hat u) \tilde{f}_1^{hard} 
- \hat s \hat u \tilde{f}_3 \right]}{(s - u)} 
\sum_q e_q^2 \, \left( H^q - \tau E^q \right), 
\nonumber \\
C &\equiv& \int_{-1}^1 \frac{dx}{x} \, \tilde{f}_1^{hard} \, 
\mathrm{sgn}(x) \, \sum_q e_q^2 \, \tilde H^q, 
\nonumber 
\end{eqnarray}
To estimate the amplitudes of 
Eqs.~(\ref{eq:GMhandbag})-(\ref{eq:F3handbag}), 
we need to specify a model for the GPDs. 
Following Refs.~\cite{Rad98,Die99}, we use an unfactorized (valence) model 
for the GPDs $H, E$, and $\tilde H$, in terms of a forward 
parton distribution and a gaussian factor in $x$ and $q^2$ 
(see Eq.~(68) in second Ref.~\cite{Die99} with transverse 
size parameter $a = 0.8$ GeV$^{-1}$). 
\newline
\indent
In Fig.~\ref{fig:cross}, we display the effect of $2 \gamma$ exchange on 
the cross sections.
For the form factors, we use the $G_{E p} / G_{M p}$ 
ratio as extracted from the 
polarization transfer experiments~\cite{Gayou02}. 
For $G_{M p}$, we adopt the parametrization of~\cite{Bra02} 
(scaled by a factor 0.995, as discussed further on). 
Fig.~\ref{fig:cross} illustrates that the values of $G_{E p}$ as extracted 
from the polarization data are inconsistent with the slopes 
one extracts from a linear fit to the Rosenbluth data 
in the $Q^2$ range where data from both methods exist. 
By adding the $2 \gamma$ exchange correction, using the GPD  
calculation as described above, one firstly observes that 
the Rosenbluth plot becomes slightly non-linear, in particular at the 
largest $\varepsilon$ values. Furthermore, one sees that 
over most of the $\varepsilon$ range, the slope is indeed steeper  
in agreement with the Rosenbluth data. Thirdly, 
in order to fit the data when including the $2 \gamma$ exchange 
correction one has to slightly decrease the 
value of $G_{M p}$ of~\cite{Bra02} (by a factor 0.995). 
This change in $G_{M p}$ is just the simplest estimate of how the 
additional radiative corrections would affect the extraction of 
$G_{M p}$ from the data. 
We see that including the $2 \gamma$ exchange allows to reconcile 
both polarization transfer and Rosenbluth data. 
It is clearly worthwhile to do a global re-analysis of all 
large $Q^2$ elastic data 
including the $2 \gamma$ correction, which is however 
beyond the scope of this letter. Such a re-analysis will allow one  
to relate the intercept and slopes of the dotted curves 
in Fig.~\ref{fig:cross} (obtained from the solid curves by turning off the 
$2 \gamma$ correction) to the true values of $G_{E p}$ and 
$G_{M p}$ respectively. 
\newline
\indent
The real part of the $2 \gamma$ exchange amplitude can be accessed directly 
as the deviation from unity of the ratio of $e^+ / e^-$ 
elastic scattering. Our calculation gives an $e^+ / e^-$ ratio 
of about 0.98 in the range $Q^2 = 2 - 5$~GeV$^2$ and  
at large $\varepsilon$ values, about $1 \sigma$ below the 
SLAC data~\cite{Mar68}. At smaller values of $\varepsilon$, 
where no data exist at moderately large $Q^2$, 
our prediction for the $e^+/e^-$ ratio rises 
and becomes larger than 1 around $\varepsilon = 0.35$. 
It will be an important test of the calculation to check this trend. 
\newline
\indent
A test of the imaginary part of the $2 \gamma$ exchange amplitudes is 
obtained using $A_n$, shown in Fig.~\ref{fig:an}. 
Our partonic estimate for $A_n$ displays a forward peaking, and 
reaches a significant part of 1 \%. It may provide a 
useful cross-check of the $2 \gamma$ amplitude.  
\newline
\indent
In summary, we have estimated the two-photon exchange contribution to 
elastic $ep$ scattering at large $Q^2$ in a partonic calculation, and 
were able to express this amplitude in terms of the GPDs of the nucleon. 
We found that the $2 \gamma$ exchange contribution is able to quantitatively 
resolve the existing discrepancy between Rosenbluth and polarization transfer 
experiments.
\newline
\indent
This work was supported by the Taiwanese NSC under contract 
92-2112-M002-049 (Y.C.C.), 
by the NSF under grant PHY-0245056 (C.E.C.)
and by the U.S. DOE under contracts DE-AC05-84ER40150 (A.A., M.V.) 
and DE-AC03-76SF00515 (S.J.B.). 
M.V. also thanks P. Guichon for helpful discussions. 

\end{document}